\documentclass{webofc}
\usepackage[labelfont=bf]{caption} 
\usepackage[varg]{txfonts}   
\usepackage{hyperref}
\usepackage{url}

\usepackage{siunitx}
\hypersetup{colorlinks=true,citecolor=blue,urlcolor=blue,linkcolor=blue}
\usepackage{subcaption}
\RequirePackage[labelformat=simple]{subcaption}

\usepackage{ragged2e}
\usepackage{placeins}

\usepackage{tikz}
\usepackage{tikz-3dplot}
\usepackage{pgfplots}
\pgfplotsset{compat=1.18}
\usetikzlibrary{positioning,shapes,decorations.markings,arrows.meta,decorations.pathmorphing,patterns}
\newcommand{\xmark}[1]{\textcolor{red}{\ding{55}}}%

\usepackage{bbding}
\newcommand{\MyArrow}[4]{
        \draw[line width=0.5pt] (#1,#2) -- (#3,#4); 
        \draw[-stealth, line width=0.5pt](#1,#2) -- (#1+#3/2-#1/2,#2+#4/2-#2/2);
}
\newcommand*{\Reg}{\ensuremath{\mathbb{R}}\xspace}
\newcommand*{\Pom}{\ensuremath{\mathbb{P}}\xspace}
\begin{document}
\title{Light-Meson Spectroscopy at COMPASS}
%
%

\author{\firstname{David} \lastname{Spülbeck}\inst{1}\fnsep\thanks{\email{david.benjamin.spulbeck@cern.ch}} \\
        for the COMPASS Collaboration}

\institute{Helmholtz-Institut für Strahlen- und Kernphysik, Universität Bonn, 53115 Bonn, Germany 
}

\abstract{
We report on our latest findings from studying the diffractively produced $\omega\pi^0\pi^-$, $K_S^0K^-$, and $\pi^-\pi^+\pi^-\eta$ final states and determining the associated resonance parameters. A particular focus is on the signal of the lightest hybrid meson candidate with spin-exotic quantum numbers $J^{PC}=1^{-+}$, i.e. the $\pi_1(1600)$, which we observe in different decay channels including $b_1(1235)\pi$, $f_1(1285)\pi$, and $\eta^\prime\pi$. Furthermore, we highlight the observation of a supernumerary state with respect to predictions of quark models in the strange sector with $J^P=0^-$.
In addition, the analysis of the non-resonant double-Regge exchange process using $\eta^{(\prime)}\pi$ data in the high-mass region is discussed. 
It is the first event-based likelihood fit of a double-Regge model and improves our understanding of non-resonant production mechanisms in light-meson spectroscopy.
}
\maketitle
\setlength{\footskip}{3.60004pt}
\section{Introduction}
\label{intro}
Quantum Chromodynamics (QCD) is widely accepted as the fundamental theory of the strong interaction and as such should also explain how quarks and gluons form hadrons.
At high energies QCD can be solved by perturbation theory, and its predictions are in precise agreement with experimental observations. 
However, at low energies, which correspond to the length scale of hadrons, the perturbative approach fails because of the running coupling of the strong interaction. 
In this regime, effective theories are used to make quantitative predictions. One popular approach to test these models is to study the excitation spectrum of hadrons. Here, the predictions are compared to experimental measurements to gain insight into the underlying mechanisms of QCD. 
The simplest effective model, i.e. the Constituent Quark Model (CQM), is already successful in describing the excitation spectrum in the light-quark sector in many cases. In this model, baryons consist of three constituent quarks ($q\,q^\prime q^{\prime\prime}$) or three antiquarks ($\bar{q}\,\bar{q}^\prime \bar{q}^{\prime\prime}$), whereas mesons consist of a quark-antiquark pair ($q\bar{q}^\prime$). 
The latter are characterised by the quantum numbers $J^{PC}$, where $J$ is the total spin of the system, $P$ is its parity which depends on the relative orbital angular momentum $L$ between the two quarks, and $C$ is its charge conjugation parity, which is a good quantum number only for states made of a quark and its own antiquark. 
In the CQM, mesons cannot have certain combinations of quantum numbers, i.e. $J^{PC}=0^{--},(\text{even})^{+-},(\text{odd})^{-+}$, which are called spin-exotic configurations. 
Hence, any meson with such quantum numbers has configurations beyond the CQM and is therefore exotic.
In the past years significant progress has been achieved in Lattice QCD (LQCD), the numerical solution of QCD on a discrete space-time lattice, which is in addition often coupled with effective theories. 

The COMPASS experiment, an approximately \SI{60}{m} long multi-purpose two-stage forward spectrometer, recorded the world's largest data set of diffractive pion and kaon dissociation, which allows precision measurements of the light-meson spectrum. At a beam momentum of \SI{190}{GeV/\textit{c}}, the hadron beam provided by CERN's M2 beam line consists of 96.8~\% pions and 2.4~\% kaons. 
COMPASS measured events where beam particles scatter off a fixed proton target at low reduced four-momentum transfers\footnote{$t^\prime=\left|t\right|-\left|t\right|_\text{min}\approx \left|t\right|$}, \SI{0.1}{(GeV/\textit{c})^2} $<t^\prime<$ \SI{1.0}{(GeV/\textit{c})^2}. 
This scattering process is described by the $t$-channel exchange of a single Reggeon ($\Reg$), which in this kinematic regime is predominantly Pomeron ($\Pom$) exchange. 
Interacting with the Pomeron, the beam meson is excited to a state $X$ (resonance production) which decays to a final state of hadrons ($h_1, h_2, .., h_n$). 
This is schematically depicted in Fig.~\ref{fig:prod} (left). 
The state $X$ is specified by the quantum numbers $J^{PC}M^\varepsilon$, where $M$ is the projection of the total spin $J$ along the quantisation axis, and $\varepsilon$ is the reflectivity, which  coincides with the naturality of the exchanged Reggeon ($\varepsilon=+$ for Pomeron exchange) at these energies. 
The scattering process can be fully described by the variables $m_X$, $t^\prime$, and $\tau_n$, where the latter symbol denotes the kinematic variables of the $n$-body decay of $X$. 
At higher masses of $X$ approximately $m_X\gtrsim\SI{3.0}{GeV/\textit{c}^2}$, multi-Regge exchanges start to dominate. 
The analysis of this kinematic regime is interesting for several reasons: on the one hand, the applicability of Regge theory can be tested and the dominant exchanges can be extracted. On the other hand, a better understanding of the high-mass regime will allow us to constrain non-resonant contributions at lower masses.
The production of $\eta^{(\prime)}\pi$ through the double Regge event topology (double Regge production) is depicted in Fig.~\ref{fig:prod} (right). 
The scattering process is fully described by the kinematic variables $t_{\eta^{(\prime)}/\pi}$, $t_p$, $s_{\eta^{(\prime)}\pi}$, and $s_{\eta^{(\prime)}/\pi p}$, as indicated in Fig.~\ref{fig:prod} (right).

\newcommand{\DrawDiffProd}{
\begin{tikzpicture}[scale=0.9, transform shape]

\MyArrow{-1.5}{0.5}{0}{0};
\node[align=left] at (-2.3,0.5) {$(\pi^-,K^-)$};

\draw[{-stealth}, line width=0.5pt](0,0) -- (0.5,0);
\draw[line width=0.5pt] (0,0) -- (1,0);
\node[align=left] at (0.5,0.5) {$X^- [J^{PC}M^{\varepsilon}]$};

\MyArrow{1}{0}{3}{0.7};
\node[align=left] at (3.3,0.7) {$h_1$};
\MyArrow{1}{0}{3}{0.35};
\node[align=left] at (3.3,0.35) {$h_2$};
\MyArrow{1}{0}{3}{0.0};
\node[align=left] at (3.3,0.0) {$h_3$};
\node[align=left] at (3.3,-0.35) {\tiny{$\vdots$}};
\MyArrow{1}{0}{3}{-0.7};
\node[align=left] at (3.3,-0.7) {$h_n$};

\draw[line width=0.5pt] (-0.03,-1) -- (-0.03,0);
\draw[line width=0.5pt] (0.03,-1) -- (0.03,0);
\node[align=left] at (-0.7,-0.5) {\Pom,\Reg};

\draw[gray, -To, line width=2pt](0.4,-0.2) -- (0.4,-0.8);
\node[gray, align=left] at (0.8,-0.4) {\textbf{$t^\prime$}};

\MyArrow{-1.5}{-1.5}{0}{-1};
\node[align=left] at (-2.3,-1.5) {$p_{\text{target}}$};

\MyArrow{0}{-1}{3}{-1.5};
\node[align=left] at (3.5,-1.5) {$p_{\text{recoil}}$};

\fill[black] (0,0) circle (0.5ex);
\fill[black] (0,-1) circle (0.5ex);
\fill[black] (1,0) circle (0.5ex);

\end{tikzpicture}}

\newcommand{\DrawDoubleRegge}{
\begin{tikzpicture}[scale=0.9, transform shape]
            
                        \MyArrow{-1.5}{0.5}{0}{0.5};
                        \node[align=left] at (-1.8,0.5) {$\pi^-$};
                        
                        \MyArrow{0}{0.5}{2}{0.5};
                        \node[align=left] at (2.6,0.5) {$\eta^{(\prime)}/\pi^-$};

                        \MyArrow{0}{-0.5}{2}{-0.5};
                        \node[align=left] at (2.6,-0.5) {$\pi^-/\eta^{(\prime)}$};

                        \draw[line width=0.5pt,color=black] (-0.03,-1.5) -- (-0.03,-0.5);
                        \draw[line width=0.5pt,color=black] (0.03,-1.5) -- (0.03,-0.5);
                        \node[align=left] at (-0.7,0.0) {\Reg};
            
                        \draw[line width=0.5pt,color=black] (-0.03,-0.5) -- (-0.03,0.5);
                        \draw[line width=0.5pt,color=black] (0.03,-0.5) -- (0.03,0.5);
                        \node[align=left] at (-0.7,-1.0) {\Reg};
   
                        \MyArrow{-1.5}{-1.5}{0}{-1.5};
                        \node[align=left] at (-2.0,-1.5) {$p_{\text{target}}$};
                        
                        \MyArrow{0}{-1.5}{2}{-1.5};
                        \node[align=left] at (2.5,-1.5) {$p_{\text{recoil}}$};
                        
                        \fill[black] (0,0.5) circle (0.5ex);
                        \fill[black] (0,-0.5) circle (0.5ex);
                        \fill[black] (0,-1.5) circle (0.5ex);

                        \draw[<->,thick]
                        (3.1,0.4) to[bend left=30]
                        node[midway,right] {$s_{\eta^{(\prime)}\pi}$}
                        (3.1,-0.4);  
                        \draw[<->,thick]
                        (3.1,-0.6) to[bend left=30]
                        node[midway,right] {$s_{\eta^{(\prime)}/\pi p}$}
                        (3.1,-1.4);  

                        \draw[<->,thick]
                        (1.75,-1.7) to[bend left=25]
                        node[midway,above] {$t_{p}$}
                        (-1.2,-1.7);  
                        \draw[<->,thick]
                        (-1.2,0.7) to[bend left=25]
                        node[midway,below] {$t_{\eta^{(\prime)}/\pi}$}
                        (1.8,0.7);  

                    \end{tikzpicture}
}

\begin{figure}
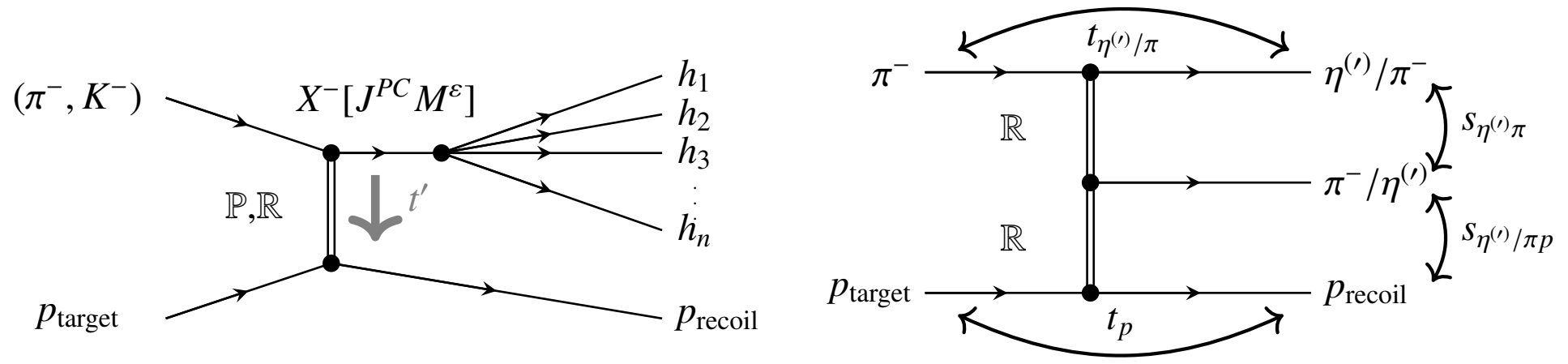

    \begin{subfigure}{0.5\linewidth}
        \centering
        \DrawDiffProd
        \vspace{0.2cm}
    \end{subfigure}
    \begin{subfigure}{0.5\linewidth}
        \centering
        \DrawDoubleRegge
    \end{subfigure}
    \caption{The production mechanisms of interest. Diffractive dissociation into a multi-hadron final state (left) and production of the $\eta^{(\prime)}\pi$ system via the exchange of two Reggeons (right).}
    \label{fig:prod}
\end{figure}


\section{Modelling the scattering amplitudes}
\label{sec:model}

%

\subsection{Resonance production}\label{sec:res_prod}
The differential cross section for resonance production is proportional to the modulus squared of the matrix element $\mathcal{M}$. We model it by the coherent sum over all partial waves $j$ using the separation Ansatz
\begin{align}
    \mathcal{M}(m_X,t^\prime,\tau_n)=\sum_j\underbrace{\mathcal{P}_\Pom(m_X,t^\prime)\cdot\mathcal{D}_j(m_X,t^\prime)}_{T_j(m_X,t^\prime)}\cdot\Psi_j(\tau_n),
\end{align} 
with the Pomeron production amplitude $\mathcal{P}_\Pom$, the propagation amplitude $\mathcal{D}$, and the decay amplitude $\Psi$. The propagation can include multiple resonances and non-resonant components. The amplitude analysis is divided into two stages. First, the data are divided into narrow kinematic cells of $m_X$ and $t^\prime$, so that one can assume that within one cell the total scattering amplitude is constant with respect to these two variables. 
For each cell, the total amplitude is decomposed into individual partial waves, which is why this stage is called partial-wave decomposition (PWD).
Each partial wave describes the decay of a potential resonance with well defined quantum numbers $J^{PC}M^{\varepsilon}$ into the final state as a sequence of two-body decays using the helicity formalism. 
Intermediate two-body resonances in this sequence are called isobars.
In extended negative log-likelihood fits, the strengths and relative phases of each partial wave are extracted, summarised by the so-called transition amplitudes $\mathcal{T}_j$, which form the spin-density matrix $\rho_{ij}=\mathcal{T}_i\mathcal{T}_j^*$. 

In the second step, the dependences of the transition amplitudes on $m_X$ and $t^\prime$ are modelled by fitting a resonance model to the transition amplitudes. It is therefore called the resonance-model fit (RMF). 
Here, Breit--Wigner amplitudes with energy-dependent widths are summed to describe the different resonances in each partial wave. 
Furthermore, for each partial wave a coherent non-resonant contribution is added, which is modelled by the following  phenomenological amplitude:
\begin{align}
D_j^\text{n.r.}(m_X,t^\prime)=\mathcal{C}_j^\text{n.r.}(t^\prime)\cdot(m_X-m_\text{thr.})^{a_j} \exp\left[-b_j\cdot q_j^2(m_X)\right].
\end{align}
$m_\text{thr.}$ is the threshold mass of the final-state system, $\mathcal{C}_a^\text{n.r.}$ is a complex-valued strength coefficient (fit parameter), $a_j$ and $b_j$ are shape parameters, and $q_j^2(m_X)$ is the two-body break-up momentum of the first two-body decay of wave $j$. More details on the amplitude analysis applied to COMPASS data can be found in Refs.~\cite{LightMesonSpec,3PiPWD,3PiRMF}.

\subsection{Double-Regge production}\label{sec:dr_prod}
The double Regge scattering amplitude is described using the model of Shimada et al. \cite{Shimada}. Each possible double-Regge configuration is associated with one amplitude $T_j$ that takes into account the Regge trajectories, parameterised by the linear function $\alpha_\Reg(t)$, as well as the invariant mass squared of the final-state particles coupling to the Reggeon, i.e. $s_i\in\{s_{\eta^{(\prime)}/\pi p}$,$s_{\eta^{(\prime)}\pi}\}$. 
We multiply the Shimada amplitude with the form factors $F_\Reg(t;b_\Reg)=\exp(b_\Reg t)$ that depend on the slope parameters $b_\Reg$, which are free fit parameters. Each amplitude has a real valued strength coefficient $c_j$, which is also a free fit parameter. The cross section is proportional to the squared matrix element, which finally reads  
\begin{align}
    \mathcal{M}(\vec{c},\vec{b},\tau)=\sum_{j\in\{(\Reg_1,\Reg_2)\}} c_j \underbrace{F_{\Reg_1}(t_{\eta^{(\prime)}/\pi},b_{\Reg_1}) F_{\Reg_2}(t_p,b_{\Reg_2})}_\text{Form factors} \underbrace{T(\alpha_{\Reg_1} (t_{\eta^{(\prime)}/\pi} ),  \alpha_{\Reg_2} (t_p); s_{i})}_\text{Shimada model},
\end{align} 
with $\tau$ symbolising the kinematic variables $\tau=\{t_{\eta^{(\prime)}/\pi},t_p,s_{\pi p},s_{\eta^{(\prime)}\pi}\}$. Note that three kinematic variables are sufficient to describe these kinds of reactions. 

\section{Results}
\subsection{Resonance production}
Building on the formalism of Sect.~\ref{sec:res_prod}, we perform full amplitude analyses of the diffractively produced final states $\omega\pi^0\pi^-$, $K_S^0K^-$, and $\pi^-\pi^+\pi^-\eta$, based on 720\,k, 430\,k, and 630\,k events, respectively. All three analyses use four $t^\prime$-bins to better separate resonant and non-resonant contributions. For $\omega\pi^0\pi^-$, the total amplitude is decomposed using four isobars, and the RMF extracts eleven resonances simultaneously in 23 selected partial waves, including, for the first time in COMPASS data, the decay $\pi_1(1600)\to b_1(1235)\pi$. For $K_S^0K^-$, selection rules allow only $a_\text{even}$ resonances, which reduces the number of partial waves; the RMF extracts six resonances in four partial waves with high statistical precision. For $\pi^-\pi^+\pi^-\eta$, 15 isobars are used, and eleven resonances are extracted simultaneously in 17 partial waves. The main RMFs yield $\chi^2/\text{n.d.f.}$ values of 6413/4572 ($\omega\pi\pi$), 2097/1246 ($K_SK$), and 5200/3264 ($\pi\pi\pi\eta$).

\begin{figure}[hb]
\centering
\sidecaption
\includegraphics[width=9cm,clip]{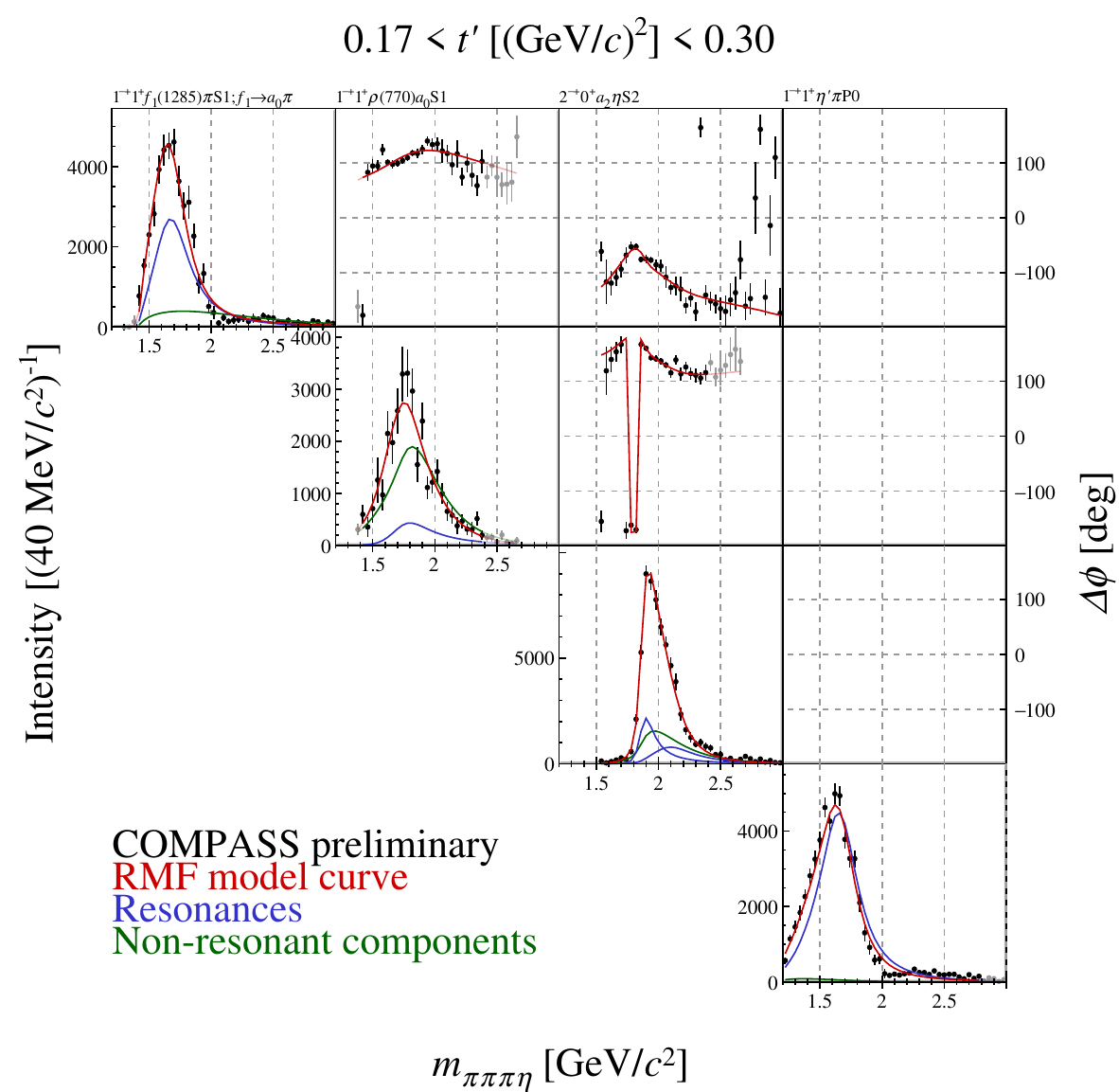}
\captionfont
\caption{\justifying Visualization of the spin-density submatrix of the $\pi\pi\pi\eta$ analysis for the three $J^{PC}=1^{-+}$-waves and one $J^{PC}=2^{-+}$ wave for reference
in the first $t^\prime$-bin. The waves are indicated by the labels in the top line. On the diagonal are the intensities and
on the upper triangle the relative phase differences. The black data points represent the spin-density matrix
elements obtained by the PWD and the curves are the total model (red) and its components (blue and green). As the $\eta^\prime$-waves are treated incoherently with respect to the others, these two kinds of waves do not have relative phases.}
\label{fig:matrixplot}       
\end{figure}

Fig.~\ref{fig:matrixplot} shows, as an example, a submatrix of the spin-density matrix extracted from the $\pi^-\pi^+\pi^-\eta$ analysis, comprising three spin-exotic $J^{PC}=1^{-+}$ waves together with the dominant $J^{PC}=2^{-+}$ wave, which is included as a phase reference. The black data points are the spin-density matrix elements obtained from the PWD. The diagonal panels show the wave intensities, while the off-diagonal panels show the relative phase between the respective pair of waves. The red curve is the total resonance model, decomposed into its individual contributions. In the $1^{-+}$ sector we observe the lightest hybrid meson candidate $\pi_1(1600)$, and its decays into $\eta^\prime\pi$, $f_1(1285)\pi$, and $\rho(770)a_0(980)$. The latter two are observed here for the first time with COMPASS data and the decay $\pi_1(1600)\to\rho(770)a_0(980)$ is observed for the first time in general. The $J^{PC}=2^{-+}$ reference wave is well described by two resonances, compatible with $\pi_2(1880)$ and $\pi_2(2100)$, and the non-resonant contribution.

Based on the RMF of the $\pi^-\pi^+\pi^-\eta$ analysis, we calculate the relative branching ratio for the decays of $\pi_1(1600)$ into $f_1(1285)\pi$ and $\eta^\prime\pi$ using the two partial waves $1^{-+}1^+f_1(1285)\pi S1;f_1\to a_0\pi$ and $1^{-+}1^+\eta^\prime\pi P0$. Each branching fraction is proportional to the integrated resonance contribution summed over all $t^\prime$ bins. The measured branching-fraction ratio is
\begin{align}
    \underbrace{\frac{\Gamma^{\pi_1(1600)}_{f_1(1285)\pi;f_1(1285)\to a_0^\pm\pi^\mp\to \eta\pi^-\pi^+}}{\Gamma^{\pi_1(1600)}_{\eta^\prime\pi;\eta^\prime\to \eta\pi^-\pi^+}}}_{\text{BR}_\text{meas.}}
    = \underbrace{\frac{\Gamma^{\pi_1(1600)}_{f_1(1285)\pi}}{\Gamma^{\pi_1(1600)}_{\eta^\prime\pi}}}_{\text{BR}_\text{red.}} \cdot
    \underbrace{\frac{\Gamma^{\eta^\prime}_{\eta\pi^+\pi^-}}{\Gamma^{f_1(1285)}_{ a_0^\pm(980)\pi^\mp}\Gamma^{a_0^\pm(980)}_{\pi^\pm\eta}}}_\text{PDG}=0.605 \pm 0.020^{+0.101}_{-0.027}\, ,
\end{align}
where the first uncertainty is the statistical uncertainty and the second one is the systematic uncertainty, determined by varying the model and taking the maximum observed deviation as the uncertainty, as done in \cite{3PiRMF}. See \cite{haas_phd,beckers_phd} for further details. Accounting for the branching fractions of the isobars involved, which we take from the PDG \cite{PDG}, and accounting for the unobserved neutral channels, we obtain a reduced branching-fraction ratio of $\text{BR}_\text{red.}=1.01 \pm 0.03^{+0.17}_{-0.04}\pm 0.11$, where the last uncertainty is the propagation of the uncertainties of the PDG branching fractions\footnote{The upper systematic deviation arises from relaxing the event-selection cuts, while the lower deviation results from using the mean of 2000 bootstrap samples to estimate the transition amplitudes instead of the likelihood-based estimates from the nominal PWD fit.}. Effective models like the flux-tube model \cite{FTM} and first LQCD calculations \cite{LQCD} predict a higher ratio for a hybrid meson. Definitive conclusions should be drawn only once more accurate predictions become available. In addition, reaching final conclusions requires determining this branching ratio for the $b_1(1235)\pi$ channel, since it is expected to be the dominant decay mode. Here an accurate measurement of the branching $b_1(1235)\to\omega\pi$ is needed.

\begin{figure}[hb]
\centering
\includegraphics[width=\textwidth]{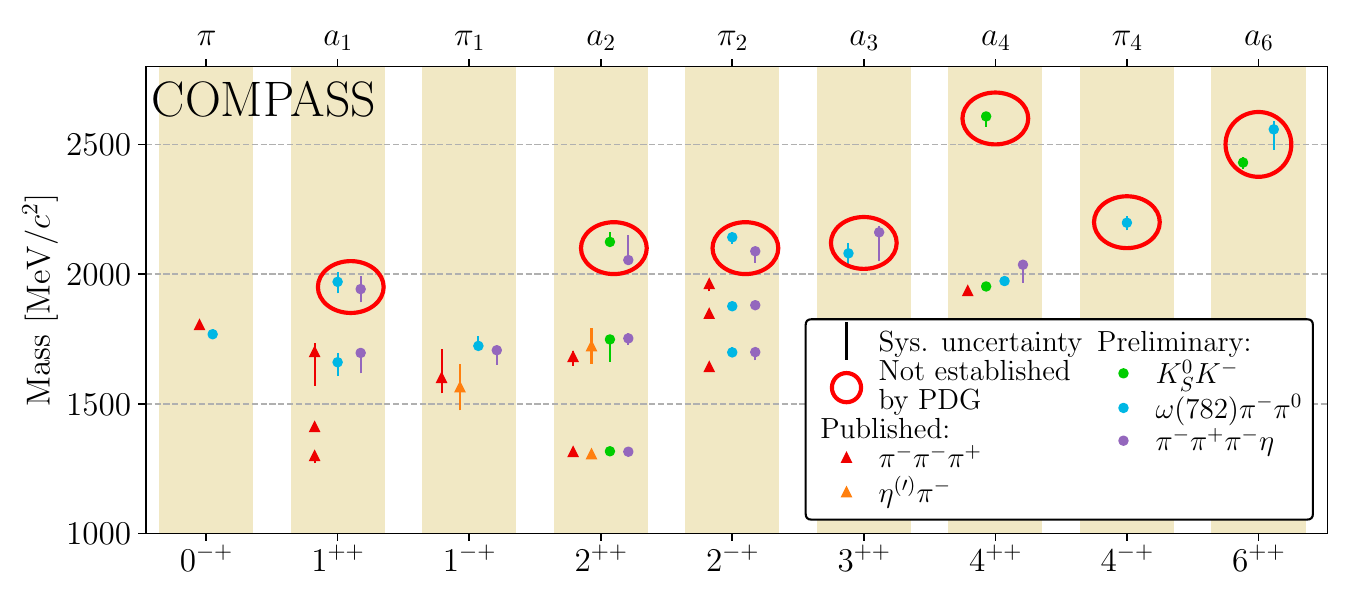}
\caption{Light isovector excitation spectrum as measured in six different final states with COMPASS data. The results of $\pi^-\pi^+\pi^-$ and $\eta^{(\prime)}\pi$ are published in Refs.~\cite{3PiRMF,PiOneJPAC}.}
\label{fig:isovec_spec}       
\end{figure}

Combining the results of these three analyses with the previously published measurements of $\pi^-\pi^+\pi^-$~\cite{3PiRMF} and $\eta^{(\prime)}\pi$~\cite{PiOneJPAC}, Fig.~\ref{fig:isovec_spec} gives an overview of the light isovector meson spectrum as observed by COMPASS in six different final states. Despite being obtained from independent analyses, a remarkably consistent picture emerges. We find clear evidence for seven states not yet established by the PDG, most of them above \SI{2}{GeV/\textit{c}^2}. The systematic uncertainties exceed the statistical ones for most resonances, highlighting the need for more sophisticated, unitary resonance models for multi-body final states.

Using the kaon component of the beam, we further analysed the $K^-\pi^-\pi^+$ final state with the same strategy, based on 730\,000 events \cite{SW}. Limited particle-identification acceptance creates holes that reduce sensitivity in part of the kinematic region and partial waves. Nevertheless, eleven resonances are extracted, shedding new light on the strange light-meson sector. The highlight is the observation of the $K(1690)$ with quantum numbers $J^P=0^-$ at high statistical significance. It is the third state observed in this sector, while the CQM predicts only two \cite{StrangeCQM}, which makes it a direct candidate for a supernumerary meson in the strange sector.

\vspace{-0.5cm}
\subsection{Double Regge}

\begin{figure}[h]
    \centering
    \hbox{
     \begin{subfigure}{0.32\linewidth}
        \centering
        \includegraphics[width=1.15\textwidth,trim=4mm 0 8mm 0, clip]{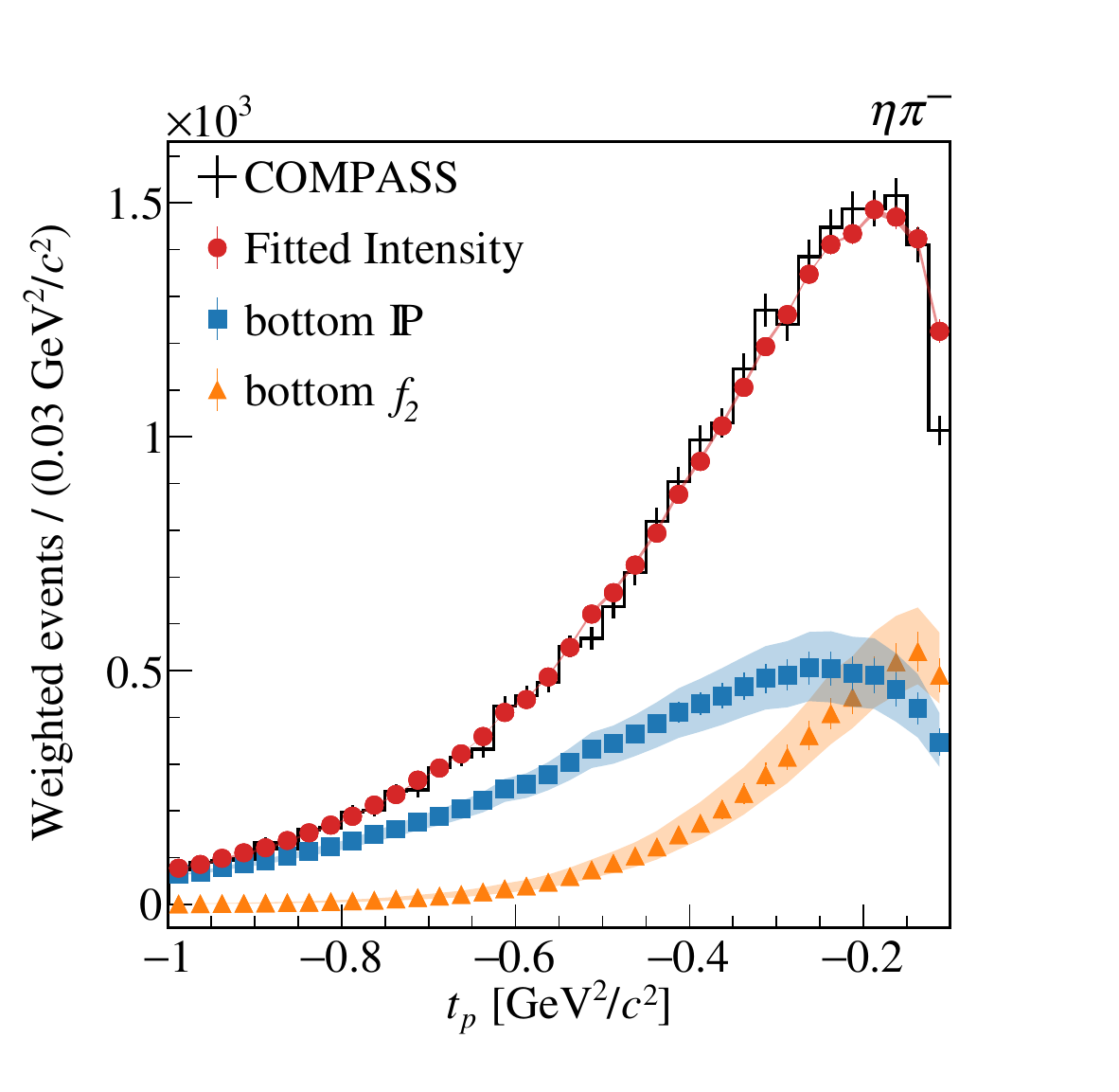}
    \end{subfigure}
    \begin{subfigure}{0.32\linewidth}
        \centering
        \includegraphics[width=1.15\textwidth,trim=4mm 0 8mm 0, clip]{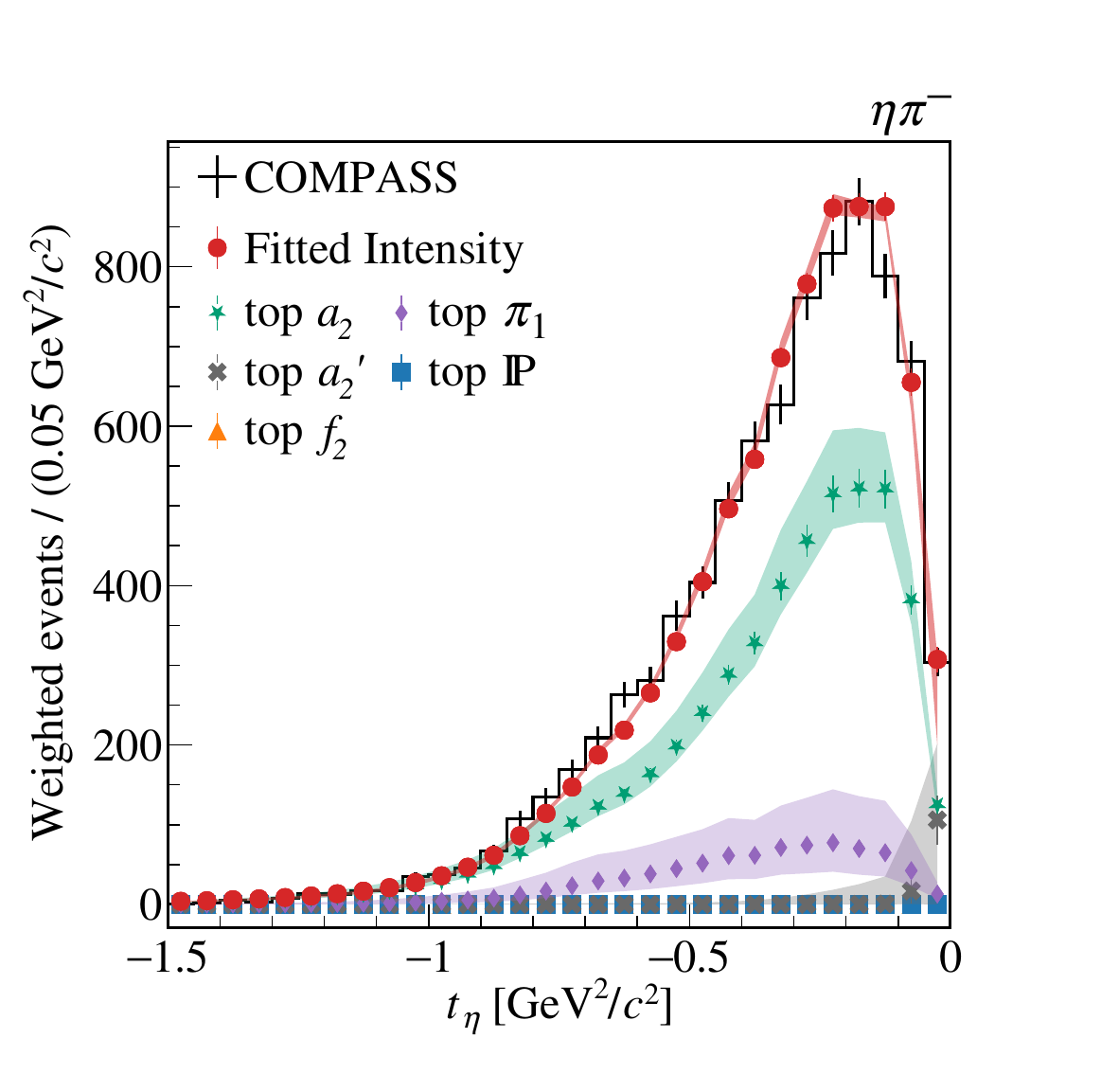}
    \end{subfigure}
    \begin{subfigure}{0.32\linewidth}
        \centering
        \includegraphics[width=1.15\textwidth,trim=4mm 0 8mm 0, clip]{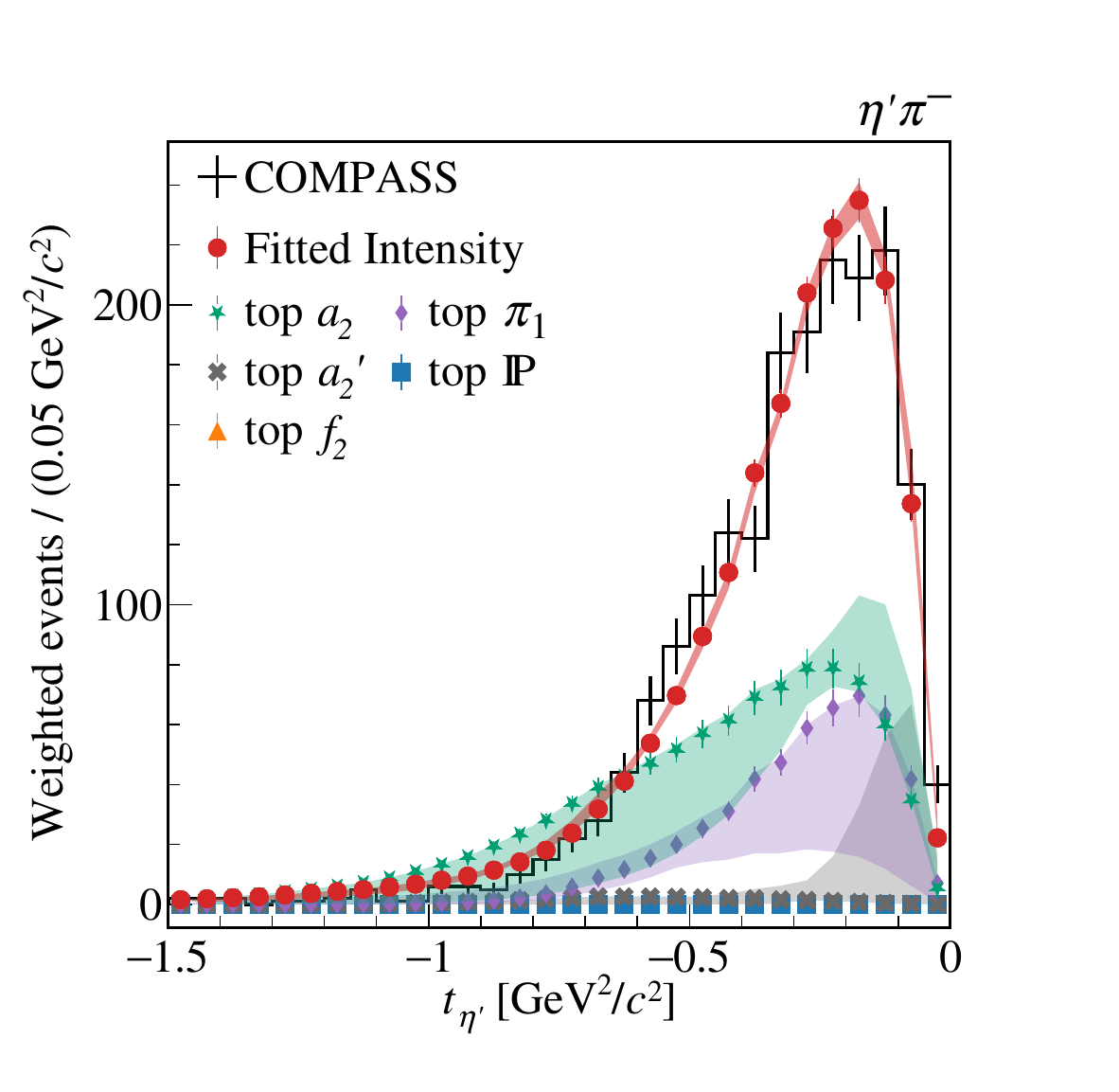}
    \end{subfigure}
    }
    \caption{Distribution of $t_p$ for $\eta\pi$ (left), showing the dominant bottom-Pomeron contribution. Distribution of $t_\eta$ for $\eta\pi$ (centre), showing the significant top-$\pi_1$ contribution. Distribution of $t_{\eta^\prime}$ for $\eta^\prime\pi$ (right), showing the significant top-$\pi_1$ contribution.}
    \label{fig:dr_results}
\end{figure}

Employing the double-Regge model introduced in Sect.~\ref{sec:dr_prod}, we carry out, together with colleagues from the Joint Physics Analysis Center (JPAC), the first unbinned extended maximum-likelihood fit to the full COMPASS $\eta^{(\prime)}\pi$ data set in the high-mass region \cite{DR_COMPASS_JPAC}. Fig.~\ref{fig:dr_results} shows one-dimensional projections of the fit result onto the relevant kinematic variables, comparing the data (black histogram) to phase-space Monte Carlo events that are propagated through the detector acceptance and reweighted according to the fitted model (red markers). The individual contributions of the exchanged-Reggeon combinations are shown separately, which allows the dominant trajectories to be identified.

Fig.~\ref{fig:dr_results} (left) shows the $t_p$ distribution for the $\eta\pi$ final state, with the uncertainty band given by the systematic model change with the highest deviation from the main model. The fit describes the data precisely and confirms the expected dominance of the bottom-Pomeron amplitude, i.e.\ Pomeron exchange at the proton vertex, while towards lower values of $|t_p|$ the bottom-$f_2$ amplitude becomes increasingly important and eventually of comparable size.

The distributions of $t_{\eta^{(\prime)}}$ for the $\eta\pi$ and $\eta^\prime\pi$ final states are shown in Fig.~\ref{fig:dr_results} (centre) and (right), respectively. Both require a sizeable contribution from an exchanged top-$\pi_1$ trajectory to describe the data accurately. Comparing the likelihoods with and without this component using Wilks' theorem, we find the top-$\pi_1$ contribution to be statistically significant at the level of $9.9\,(5.5)\,\sigma$ for $\eta\pi\,(\eta^\prime\pi)$, constituting the first observation of an exotic Regge trajectory.

\vspace{-0.5cm}
\section{Summary and conclusion}

Independent amplitude analyses of six final states measured at COMPASS yield a self-consistent picture of the light isovector meson spectrum.
The $\pi^-\pi^+\pi^-\eta$ analysis provides an estimate of the relative $f_1(1285)\pi$ to $\eta^\prime\pi$ branching ratio of the $\pi_1(1600)$, a key ingredient for testing its hybrid-meson interpretation; a final conclusion requires more precise predictions and an improved measurement of the $b_1(1235)$ branching fraction.
In the strange sector, we observe the supernumerary $K(1690)$.
The first event-based analysis of the double-Regge process, using the Shimada model, reveals an exotic Regge trajectory associated with the $\pi_1$, which could be confirmed by analysing further final states, e.g.\ $\rho(770)\pi$.

AMBER, the successor of COMPASS, foresees a dedicated kaon-beam spectroscopy programme to clarify the poorly known strange-meson spectrum.
\newpage

%
%
%

\end{document}